\documentclass[aip,reprint]{revtex4-1}
\usepackage{amsmath}
\usepackage{amssymb}
\usepackage{graphicx}% Include figure files
\usepackage{dcolumn}% Align table columns on decimal point
\usepackage{bm}% bold math
\usepackage{multirow}

\draft % marks overfull lines with a black rule on the right

\begin{document}

\title{Symmetry-specific orientational order parameters for complex structures}
\author{Jack A. Logan}
\affiliation{Department of Physics and Astronomy, Stony Brook University, Stony Brook, NY 11794}
\affiliation{Center for Functional Nanomaterials, Brookhaven National Laboratory, Upton, NY 11973}
\author{Srinivas Mushnoori}
\affiliation{Department of Chemical and Biochemical Engineering, Rutgers, The State University of New Jersey, Piscataway, NJ 08854}
\author{Meenakshi Dutt}
\affiliation{Department of Chemical and Biochemical Engineering, Rutgers, The State University of New Jersey, Piscataway, NJ 08854}
\author{Alexei V. Tkachenko}
\email{oleksiyt@bnl.gov}
\affiliation{Center for Functional Nanomaterials, Brookhaven National Laboratory, Upton, NY 11973}

\begin{abstract}
A comprehensive framework of characterizing complex self-assembled structures with a set of orientational order parameters is presented. It is especially relevant in the context of using anisotropic building blocks with various symmetries. Two classes of tensor order parameters are associated with polyhedral nematic, and bond orientational order, respectively. For the latter, a variation of classical bond order parameters is introduced that takes advantage of  the symmetry of constituent particles, and/or expected crystalline phases. These Symmetrized Bond Order Parameters (SymBOPs) can be averaged over an entire system, or assigned locally to an individual bond. By combining that with bond percolation procedure, one is able to identify coherent domains within a self-assembled structure.  As a demonstration of the proposed framework, we apply it to a simulated hybrid system that combines isotropic and patchy particles with octahedral symmetry. Not only does the methodology allow one to identify individual crystalline domains, but  it also detects coherent clusters of a peculiar compact amorphous structure that is not space-filling and lacks any long-range order.          
\end{abstract}

\maketitle

\section{Introduction}

The concept of symmetry breaking is one of the cornerstones of modern physics, with applications ranging from fundamental interactions and the early Universe to the statistical mechanics of soft and hard  condensed matter \cite{Goldstone1961,Anderson,Chaikin1995,degennes1995}. In most cases, it may be well characterized by  the appropriate choice of {\it  order parameter} whose form  is dictated by the symmetry itself. This allows  a unified conceptual framework to be applied  in very different contexts for the description of phase transitions, topological defects, soft modes, etc. Ideally, an order parameter is a mathematical object that  is (i) zero in the disordered phase, and (ii) invariant under the remaining symmetries of the ordered phase. Because of that, in addition to its magnitude, an order parameter typically has  ``phase" variable(s). They  account for the continuous degrees of freedom along which the symmetry is broken, and hence for the Goldstone soft modes \cite{Goldstone1961,Anderson} around the emerged symmetry-broken ground state.   

It is ironic that, while much of our intuition about  symmetry-broken states of  matter comes from   crystals,  the very  definition of an order parameter in that case is highly non-trivial. On the one hand, a perfect crystal is characterized by Bragg diffraction peaks in its structure factor. This  makes their amplitudes  natural candidates as components of the order parameter. Indeed,  crystallization  can be viewed as a result of the instability of the liquid phase with respect to density fluctuations at  finite wave vectors \cite{Alexander_PRL,Braz_75, Tanaka2012}. 
However, breaking of the translational symmetry in crystals leads to breaking of the rotational symmetry as well. The opposite is not always true: there are well known  examples of phases with broken rotational symmetries that  remain spatially uniform:   nematic liquid crystals, and the 2D  hexatic phase \cite{Nelson_2D}. In fact these two cases represent two distinct ways in which orientational order may emerge: from long-range correlations in orientations of constituent particles, or from similar correlations between directions of interparticle bonds. The first type of ordering may be characterized by a family of  Polyhedral Nematic Order Parameters (PNOPs) \cite{polyhedralPRX,polyhedral_PRE2016,polyhedral_PRE2018,Akbari_Glotzer_2015}, which is a generalization of nematic for particles of higher symmetry in 3D (by analogy, we will refer to this kind of ordering  in 2D as  polygonal nematic). The other type of order is characterized by  Bond-Orientational Order Parameters, or simply Bond Order Parameters  (BOPs). 

The concept of bond-orientational order is key for a  fundamental understanding  of crystals, as well as of quasi-crystals and structural glasses.  It emerged in the 1970s, in the context of 2D crystallization \cite{Nelson_2D,Nelson_PRB}, motivated by the discovery of the Berezinskii-Kosterlitz-Thoules (BKT) topological phase transition \cite{Berez_71,Kosterlitz1973}. The local BOP in that case was defined as a complex number $e^{il\phi_b}$ assigned to each interparticle bond, where $\phi_b$ is the polar angle corresponding to the  direction of that bond. This complex-valued field captures the l-fold symmetry of e.g.  hexatic or hexagonal lattices (for $l=6$), and formally  resembles  order parameters in other classical models (such as the $XY$ model, superfluidity, superconductivity,  etc). A natural 3D generalization  of that order parameter was later introduced for the description of ordered structures  with  cubic symmetry \cite{Hess1980,MITUS_81,MITUS_82, Nelson_Toner}. 

The classical paper by Steinhardt et al. \cite{Steinhardt} proposed the use of leading-order scalars (rotational invariants)  constructed out of BOPs  for the description of the local structure and long-range correlations  in supercooled liquids and glasses. The focus of that paper was on icosahedral glass, thought at that time to be the 3D analogue of a hexatic phase, i.e., example of a long-range bond orientation order without crystallinity. That original picture has not found enough confirmation in later studies, but ultimately led to the discovery and theoretical understanding of a completely new state of matter, quasicrystals \cite{quasi}.    Just like the hypothetical icosahedral glass, they are characterized by bond orientation order with symmetries that are incompatible with periodic crystal lattices. However, nature found a solution which is not completely   translationally-invariant: the quasi-periodic  structures, characterized by a self-similar pattern of  peaks in their  structure factor.

The influence of Ref.~\onlinecite{Steinhardt}  was such, that the very term ``bond-orientational order parameter" is now primarily associated with the families of rotational invariants of the second and third order, known as  $Q_l$, and $W_l$, respectively.   While these parameters are indeed very good  structure descriptors, they are not full order parameters in the classical sense. In particular, they do not carry explicit information about the local symmetry of the structure, and the  ``phase" information (related to  orientation itself) is discarded. In recent decades, the  BOPs have been widely accepted  as measures of  the local degree of crystallinity, especially in computer simulations. Because of the limitations of the scalar  descriptors mentioned above, their  local versions take into account the short-range correlations of the full tensor BOPs \cite{Frenkel1996,Frenkel2004, BOP2008,BOP2018}. This enables one to  classify  the  particles in a structure, based on their neighborhood,   as ``crystal-like" or ``amorphous-like", and even  to identify local structural motifs such as e.g.  ``BCC"-like or ``FCC"-like.

 BOPs are commonly used for the characterization of nanoparticle  and colloidal  self-assembly \cite{Vasilyev_1,Vasilyev_chrom}.  An additional level of complexity in the context of modern self-assembled structures is introduced by the use of  ever more sophisticated building blocks, sometimes called ``designer particles" \cite{Glotzer_Solomon2007,Pine_patchy_2003,Oleg_hybrid_2016,Oleg_Octahedra_2020,Scior_Nat_phys}. As we have mentioned earlier, the correlations in particle orientations may be associated with  long-range polyhedral (or polygonal in 2D case) nematic order  \cite{Akbari_Glotzer_2015,polyhedral_PRE2016,polyhedral_PRE2018,polyhedralPRX}. Relatively well known examples of this type of  ordering are uniaxial and biaxial nematics, cubatic and tetratic phases \cite{Torq_superballs2010,cubatic_tensor}. This order   is   characterized by the tensor order parameter ${\widehat S}^{(l)}$, of appropriate order $l$,   and it is distinct from bond orientation order. The two orientational order parameters are likely to be coupled, but in a general case do not even need to be of the same symmetry.   
 
%  Note that  the lowest orders of  non-vanishing tensors compatible with tetrahedral, octahedral (cubic), and icosahedral symmetries are, $l=3,4$, and $6$, respectively \cite{polyhedralPRX}. 

In this paper, we revisit the original version  of tensor BOPs, and further expand their definition, making them  symmetry-specific.  Our objective is two-fold:  developing a more coherent picture of various types of order, essential in the context of complex self-assembly, and designing  better  characterization tools  for both experiments and computer simulations.

\section{Orientational order of  bonds and particles}
\subsection{Tensor Bond Order Parameters (BOPs)}
Let $\vec{r}_{ij}=\vec{r}_{j}-\vec{r}_i$ be the relative position of two particles, $i$ and $j$, in a certain structure. We assign to this pair  of particles a "bond director", the unit vector in the direction of   $\vec{r}_{ij}$: $\hat{\bf b}_{ij}=\vec{r}_{ij}/r_{ij}$, and a distance-dependent weighting factor $w_{ij}=w(r_{ij})$, which determines whether the particles are counted as neighbors. Now  the neighborhood of particle $i$  may be characterized with a set of  traceless multipoles \cite{tensors1998, Multipoles,tensors2018} associated with its bond directors:  
\begin{equation}
\label{multi}
\widehat{ \bf q}^{(l)}_i=\frac{\Lambda_l}{Z_i}\sum_j{w_{ij}\widehat{\mathcal  D} \hat{\bf b}_{ij}^{\otimes l}}
\end{equation}
Here $Z_i=\sum_j w_{ij}$ is the weighted coordination number of  particle $i$, and   $\hat{\bf b}_{ij}^{\otimes l}\equiv{b}_{ij}^\nu....{b}_{ij}^\mu$.  The normalization coefficient  that  corresponds to the traditional definition of multipoles in 3D is $\Lambda_l=(2l-1)!!/l!$.    $\widehat{\mathcal  D}$ is the ``detracing" operator \cite{Multipoles,tensors1998} that projects an arbitrary symmetric tensor of order $l$ onto a sub-space of traceless (irreducible)  tensors of that order.   For example,  for $l=2$   in $d$ dimensions, $ D_{\nu\mu}^{\nu'\mu'}=\delta_\nu^{\nu'} \delta_\mu^{\mu'}- \delta^{\nu'\mu'}\delta_{\nu\mu}/d$, so  $\widehat{\mathcal  D} \hat{\bf b}^{\otimes 2}= \hat{\bf b}^{\otimes 2}- \hat{\bf I}/d$.  
% The orthogonality relationship in  tensor and spherical  representations are  equivalent, but the inner products differ by normalization factor $\Lambda_l$ \cite{Multipoles}:
% \begin{equation}
%   \left( Q^{(l)}|Q'^{(l)}\right) =\Lambda_l^{-1}\widehat{ \bf Q}^{(l)}\cdot \widehat{ \bf Q}'^{(l)} 
% \end{equation}
The tensor BOPs for a system, or  a particular region of it, is obtained by appropriate averaging of the bond multipoles:    $\widehat{\bf Q}^{(l)}=\langle Z_i\widehat{\bf q}^{(l)}_i\rangle/\langle Z_i \rangle $. In order to avoid unequal counting of different bonds,   $\widehat{\bf q}^{(l)}_i$ for each particle  has  to be   weighted proportionally to the corresponding coordination number,  $Z_i$.

BOPs  are commonly defined by using spherical harmonics. This  is possible because  there is  a one-to-one linear mapping  between functions $Y_{lm}(\hat{\bf n})$  and traceless symmetric tensors in 3D space \cite{tensors1989,Multipoles,tensors2018,tensors2020}:  
\begin{align}
\label{spher_tensor1}
\widehat{\bf Y}^{(l,m)}&\equiv \sqrt{\frac{(2l+1)\Lambda_l}{4\pi}}
\int{\hat{\bf n}^{\otimes l}Y_{lm}(\hat{\bf n})d^2\hat{\bf n}}\\
\label{spher_tensor2}
Y_{lm}(\hat{\bf n})&= \sqrt{\frac{(2l+1)\Lambda_l}{4\pi}}  \hat{\bf n}^{\otimes l} \cdot  \widehat{\bf Y}^{(l,m)}
\end{align}
The tensors $\widehat{\bf Y}^{(l,m)}$ form an orthonormal   basis in the space of traceless symmetric tensors of order $l$. By projecting the tensor $\widehat{\bf q}_i^{(l)}$ into this basis, one  obtains BOP in the traditional  spherical harmonic representation \cite{Steinhardt,Frenkel1996,Frenkel2004}:
\begin{align}
 \left|q_i\right)_l=\frac{\widehat{ \bf q}_i^{(l)}\cdot \widehat{\bf Y}^{(l,m)}}{\sqrt{\Lambda_l}}= \frac{1}{Z_i}\sum_j w_{ij} \left|\hat{\bf b}_{ij}\right)_l \label{eq:Sph_harm_BOP}\\
 \left|\hat{\bf b}_{ij}\right)_l \equiv \left\{ \sqrt{\frac{4\pi}{2l+1}} Y_{lm}(\hat{\bf b}_{ij})\right\}_{m=-l..l} \label{eq:Sph_harm_BOP_bond}
\end{align}
Here we introduce bra-ket notations for the spherical harmonics with a given $l$. Note that switching from tensors to spherical harmonics  preserves orthogonality, but the inner product changes  by a constant factor:  $\left(q|q'\right)_l\equiv \sum_m{q^{(l.m)}q'^{(l.m)}}=\Lambda_l^{-1}\widehat{\bf q}^{(l)}\cdot \widehat{\bf q}'^{(l)}$. 
The classical  rotationally-invariant BOPs $Q_l$ and $W_l$ naturally arise as the  second- and third-order scalars that can be constructed out of the  tensor  $\widehat{\bf Q}^{(l)}$. In particular,   
 \begin{equation}
 Q_l=\sqrt{\left(Q|Q\right)_l}= \sqrt{ \Lambda_l^{-1}\widehat{\bf Q}^{(l)2}}
 \end{equation}
A similar mapping allows a representation of tensor BOPs   in  2D as a set of  $l$-fold symmetric complex  order parameters,  $q_l=\langle w_{ij} e^{i l\phi_{ij}}\rangle$.
 
\subsection{Polyhedral Nematic Order Parameters (PNOPs)}
To describe the orientational order associated with rotations of anisotropic constituent particles, PNOP has to be introduced \cite{polyhedralPRX}. This is a  traceless symmetric tensor, $\widehat{\bf S}^{(l)}$, similar to the case of BOP. For instance,  the minimal-order  tensors with axial ($C_{nh}/D_{nh}/D_{\infty h}$, $n>2$),  tetrahedral ($T_d$) and octahedral ($O/O_h$) symmetries are of 2nd, 3rd and 4th order, respectively \cite{polyhedral_PRE2016,polyhedralPRX,polyhedral_PRE2018, Akbari_Glotzer_2015}:
\begin{align}
\label{polyhedral1}
\widehat{\bf  s}^{(2)}_i&= \frac{3\hat{\bf x}^{\otimes 2}- \hat{\bf I}}{2}\\
\label{polyhedral2}
    \widehat{\bf  s}^{(3)}_i&=\sum_{\rm perm}\hat{\bf x}\otimes \hat{\bf y}\otimes\hat{\bf z}\\
    \label{polyhedral3}
    \widehat{\bf  s}^{(4)}_i&=\frac{5\left(\hat{\bf x}^{\otimes 4}+\hat{\bf y}^{\otimes 4}+\hat{\bf z}^{\otimes 4}\right)-\hat{\bf I}^{(4)}}{2}
\end{align}
Here $(x,y,z)$ is the coordinate system associated with particle $\alpha$; $I^{(4)}_{\nu\mu\nu'\mu'}=\delta_{\nu\mu}\delta_{\mu'\mu'}+\delta_{\nu\nu'}\delta_{\mu\mu'}+\delta_{\nu\mu'}\delta_{\nu'\mu}$.    $\sum_{\rm perm}$  represents a sum over all  non-equivalent permutations of tensor indices (to construct tensor $\widehat{\bf  S}^{(3)}_i$ that is invariant under the tetrahedral group $T$ without reflective symmetry,  the sum should include only  cyclic permutations \cite{polyhedralPRX}).  A non-zero global average of this  order parameter,  $\widehat{ \bf S}^{(l)}=\langle \widehat{ \bf s}^{(l)}_i\rangle$ is a signature of  long-range polyhedral nematic order in the system. 
Similarly to  BOPs, $\widehat{ \bf S}^{(l)}$ has  a convenient  spherical harmonic and complex-valued order parameter  representation in 3D and 2D, respectively.

% Note that in a general case, the symmetry of the particles and that of the global BOP may be different. Nevertheless, we  typically expect these two types of orientational order to be strongly  coupled.   

\subsection{Interplay of different types of order}
\begin{figure}[h!]
    \centering
    \includegraphics[width=1\columnwidth]{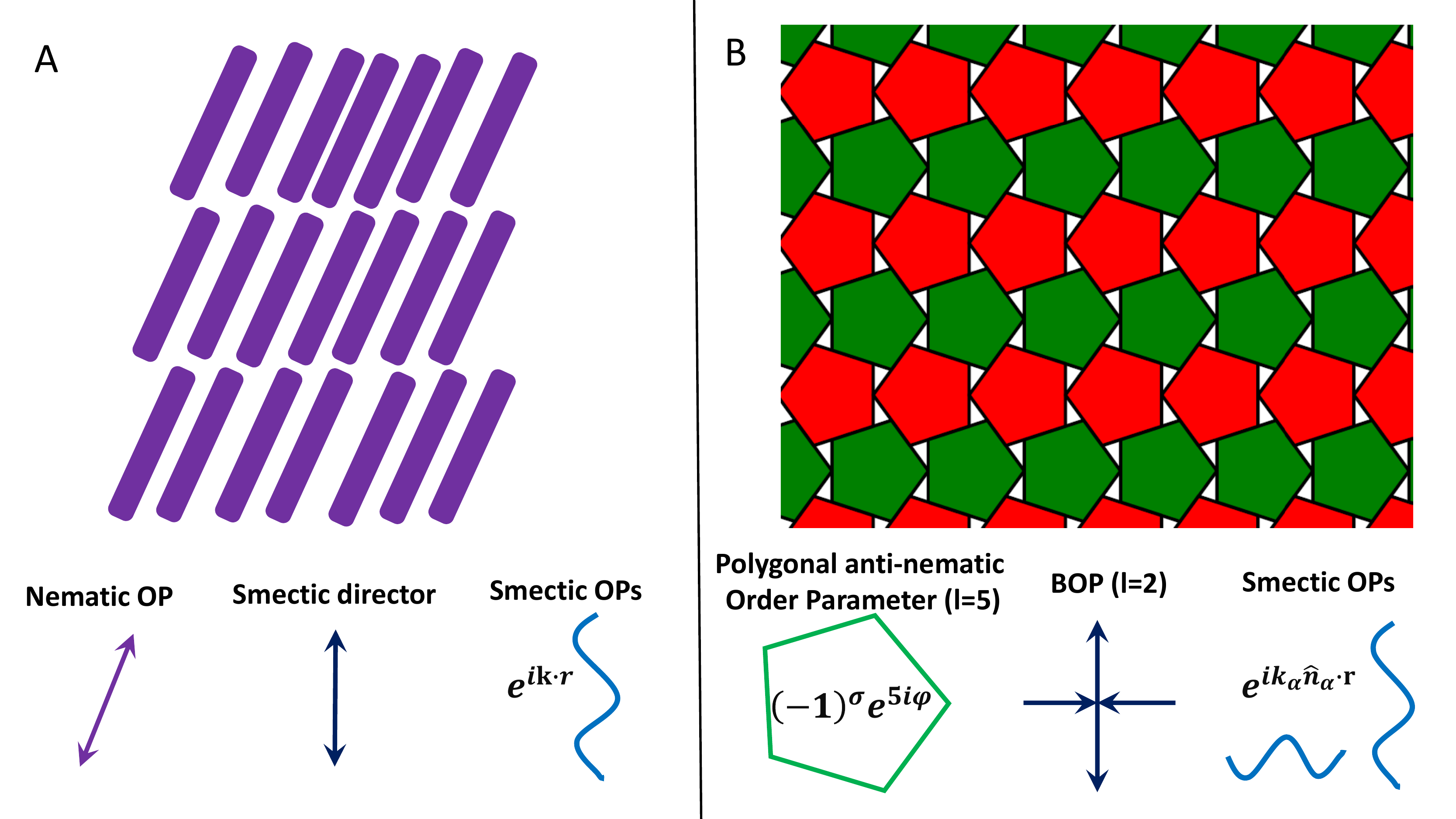}
    \caption{A: Example of generic Smectic C phase. Two second-order tensors determine nematic OP and smectic director, which are not co-aligned in this case. Spatial modulation is characterized by the complex-valued smectic OP.   B: The densest packing of regular pentagons is an example where bond-orientational order, translational order, and polygonal nematic order all coexist. The red and green colors identify the two sub-lattices ($\sigma=0,1$), respectively. Polygonal anti-nematic order is characterized by $l=5$-fold symmetric OP, the second-order BOP tensor determines principle directions of the orthorhombic lattice, and two smectic OPs characterize breaking of the translational symmetry. }
    \label{fig:pentagon_packing}
\end{figure}

By construction, BOP $\widehat{\bf Q}^{(l)}$ is a traceless symmetric tensor and, for an even order $l$,  has the same symmetry as the orthorhombic family of crystal lattices (point group $D_{2h}$). This implies that  bond orientation order in such  crystals can be described by the lowest-order non-vanishing tensor, i.e.,  quadrupole  $\widehat{ \bf Q}^{(2)}$. Its symmetry  is formally equivalent to that of a nematic tensor order parameter (uniaxial in 2D, biaxial in 3D, $d-1$ axial in $d$ dimensions). More precisely, each of the principle directions of the second-order bond orientation  tensor $\widehat{\bf  Q}^{(2)}$ plays the same role as a director of a smectic phase in liquid crystals, as illustrated in  Fig. \ref{fig:pentagon_packing}. A regular orthorhombic crystal can therefore be naturally described by a set of  $d$ de Gennes-style smectic order parameters $\psi_\alpha$ ($\alpha=1..d$), defined microscopically for each particle $i$ as \cite{degennes1995}:
\begin{align}
   \psi_{\alpha i}= e^{i\phi_\alpha(\bf{r}_i)}\\
   \nabla\phi_\alpha({\bf r})=\frac{2\pi \hat{\bf n}_\alpha}{a_\alpha}
\end{align}
Here $\phi_\alpha$ represents $d$  phase variables that change by $2\pi$ upon displacement by a lattice constant $a_\alpha$ along direction $\hat{\bf n}_\alpha$.
 Thus, the set of $d$ complex-valued smectic order parameters, together with the bond orientation quadrupole  $\widehat{ \bf Q}^{(2)}$, is a natural set of order parameters for orthorhombic crystals in $d$ dimensions. Just like with the classical order parameters, their phases correspond to degrees of freedom along which the symmetry has been broken: $d(d-1)/2$ rotations of the coordinate system, and $d$ translations that  amount to  phase shifts  $\delta \phi_\alpha$. The same correspondence holds for  the soft modes of the crystal around the ground state:  small local rotations of the bond orientation tensor are equivalent  to transverse acoustic phonons, while small  variations in phases   $\phi_\alpha$ correspond to the longitudinal ones.

The analogy to liquid crystals does not stop there. As we mentioned earlier, constituent particles themselves might not be isotropic. A general picture of a complex crystalline  assembly may be illustrated on an  example of the densest packing of pentagons, shown in Fig~\ref{fig:pentagon_packing}B. The particles are arranged  into an orthorhombic crystal. 
% which, from the point of view of bond orientation order,  corresponds to regular nematic symmetry. 
This means that  the structure is characterized by a non-zero {\it 2-fold symmetric BOP}, $q_2=\langle w(r_b)e^{2i\phi_b}\rangle$ (here $w$ is a weighting factor  which depends on the interparticle distance $r_b$).  Equivalently, this order can be described by  the second-order traceless tensor. Next, breaking of the translational symmetry can be characterized by {\it smectic} order parameters associated with each of the principle directions of the bond orientation tensor, $\hat n_\alpha$  ($\alpha=1,2$): $\psi_\alpha=\langle e^{i\hat k_\alpha {\bf n}_\alpha \cdot \bf{r}_\sigma}\rangle$, where ${\bf r}_\sigma$ are the positions of particles that belong to the same sub-lattice (e.g. $\sigma=0$ for green pentagons and  $1$ for red). $k_\alpha$ is the wavenumber for a given principle direction.  

Finally, there is orientational order of  the constituent pentagonal particles. By analogy with the  term ``polyhedral nematic" in 3D \cite{polyhedralPRX}, we will refer to this order in 2D as polygonal nematic. Well-known examples with   $l=2,4,6$-fold symmetries are  nematic, tetratic and hexatic phases, respectively.   Just like the bond-orientational order, this type of ordering in  2D can be equivalently described either by a $l$-th order tensor, or by a  complex-valued $l$-fold symmetric order parameter.    The orientational ordering  of pentagons in Fig~\ref{fig:pentagon_packing}B would correspond to breaking of $l=10$-fold symmetry. However, given that the pentagons are forming two sub-lattices, and in view of a  clear analogy to antiferromagnetic order, their arrangement  can be characterized as {\it pentagonal anti-nematic} with order parameter $S^*_5=\langle (-1)^\sigma e^{5i\phi}\rangle$ (here $\phi$ is rotation  angle for each particle, and   $\sigma =0$ or $1$,  is the index of the sub-lattice to which that particle belongs).
The  structure discussed is an example when all three types of order: bond-orientational, translational,   and polygonal (anti-) nematic, coexist.  In general, however, polygonal/polyhedral  nematic and/or bond orientation order may appear without  translational symmetry breaking. The well known  examples of that are nematic liquid crystals and the bond hexatic phase in 2D \cite{Nelson_2D}. On the other hand, random particle orientations may  co-exist with crystallinity, e.g., in  rotator phases.   

\section{\label{sec:SymBOPs} Symmetrized bond order parameters}

If the expected structure  has  a higher-order  symmetry than orthorombic, that would further constrain the space of possible  tensors $\widehat{\bf Q}^{(l)}$. 
To construct the order parameter that would automatically be consistent with that symmetry, we first introduce the notion of a  {\it reference tensor} $\widehat{\bf R}^{(l)}$. It is the $l$-th order tensor  that is  (i) invariant under all point symmetries of the expected crystal phase, (ii) traceless, and (iii) normalized to $1$, i.e.,  $|\widehat{\bf R}^{(l)}|\equiv \sqrt{\widehat{\bf R}^{(l)}\cdot \widehat{\bf R}^{(l)}}=1$. For a given reference tensor, we define the symmetrized Bond Order Parameter (SymBOP) for a bond between  particles $\alpha$ and $\beta$, as:  
\begin{equation}
\label{SymBOP}
    \widehat{\bf q}^{*(l)}_{ij}=\widehat{\bf R}^{(l)}\left(\widehat{\bf R}^{(l)} \cdot \hat{\bf b}_{ij}^{\otimes l}\right)=\widehat{\mathcal   P}^{(l)} \hat{\bf b}_{ij}^{\otimes l}
\end{equation}
Here  $\widehat {\mathcal P}^{(l)}=\widehat{\bf R}^{(l)}\otimes \widehat{\bf R}^{(l)T}$ is the projection operator ($2l$-order tensor) associated with the reference tensor $\widehat{\bf R}^{(l)}$. In the case when $\widehat{\bf R}^{(l)}$  is uniquely defined  for a given symmetry group, $\widehat{\mathcal P}$ is the projection operator associated with that group. This object is widely used, e.g., to construct a basis of molecular orbitals  consistent with a particular symmetry in quantum chemistry.  

Our definition  allows for several sub-classes  of SymBOPs, depending on how the reference tensor is chosen:
\begin{itemize}
    \item {\it Particle reference.} If the structure is built from anisotropic particles, and their local orientations are known (e.g. from computer simulations), each particle defines its own preferred  coordinate system.  The reference tensor then may be defined as $\widehat{\bf R}_i^{(l)}=\widehat{\bf s}_i^{(l)}/|{s}_i^{(l)}|$. 
     \item {\it Self-consistent reference.} In this case, tensor  $\widehat{\bf R}^{(l)}$ is  defined based on bond configurations in a particular region. The  coordinate system  $(\hat{\bf x},\hat{\bf y},\hat{\bf z})$, associated with reference tensor  $\widehat{\bf R}^{(l)}$, is rotated  to maximize the  magnitude of the SymBOP $\left|\langle \widehat{\bf Q}^{*(l)}\rangle\right|$, averaged over a certain part of the system.  In this case, the reference tensor   plays the same  role as the director of a smectic  phase. 
\end{itemize}

 Use of the auxiliary reference tensor $\widehat{ \bf R}^{(l)}$ might appear unconventional, but this is exactly what is implicitly done, even in the common case of uniaxial nematic. There, the local nematic order parameter and the director are defined based on the largest eigenvalue and the corresponding eigenvector of the second-order nematic tensor  ${\widehat S}_2$. This amounts to  applying an axially-symmetric projection operator to ${\widehat S}_2$, and maximizing the magnitude of the resulting order parameter, just like in the case of the self-consistent local reference introduced above. Since  the self-consistent  version of SymBOPs is agnostic about the orientation of the particles themselves, it may be applied  e.g., to systems of isotropic particles, or to experimental systems where information on the orientations of the individual building blocks is not available.

In the 3D case, SymBOPs can be rewritten in a spherical harmonic  representation:
\begin{equation}
\left|{ q}^*_{ij}\right)_l= \left|R \right)_l\left( R \big | \hat{\bf b}_{ij}\right)_l=\tilde{\mathcal P_l} \left|\hat{\bf b}_{ij}\right)_l
\end{equation}
Here $\left|R \right)_l$ is the  spherical harmonic version of the reference tensor: it is (i)  invariant under the given  symmetry group, and (ii) normalized so that  $\left(R \big|R \right)_l=1$. In this representation, we will refer to it as {\it reference vector}.  Table \ref{table:tensors} shows example of such vectors, for various symmetry groups. In particular, they  include spherical harmonic analogues of the invariant tensors given by Eqs. (\ref{polyhedral1})-(\ref{polyhedral3}).  

\begin{table}[h!]
\centering
\begin{tabular}{|c|c|c|} 
\hline
Symmetry & \ \ $l$\ \ \ & $\left|R\right)_l$\\
\hline
Uniaxial, e.g. $D_{\infty h}$   & 2  & $\left\{ 0,1,0\right\}$\\
\hline
Tetrahedral, $T_d$ & $3$ & $ \frac{1}{\sqrt{2}}\left\{ 0,1,0,1,0\right\}$\\
\hline
Octahedral, $O/O_h$ & $4$ & $\frac{1}{\sqrt{24}}\left\{\sqrt{5} ,0,0,0,\sqrt{14},0,0,0,\sqrt{5}\right\}$\\
& $6$ & $\frac{1}{4}\left\{0,0,\sqrt{7} ,0,0,0,-\sqrt{2},0,0,0,\sqrt{7},0,0\right\}$\\
\hline
\end{tabular}
\caption{Examples of reference vectors in the spherical harmonic representation,  for various symmetries. The coordinate system is aligned with the symmetry axes.}
\label{table:tensors}
\end{table}

The traditional BOP may be assigned to an entire system, to a certain region, or to individual particles. The latter approach is especially important for the characterization of highly heterogeneous  systems that may contain fractions with  various degrees of crystallinity.    In contrast,  SymBOPs contain non-trivial information already on the level of an individual bond. As a result,  one may identify bonds that are consistent with a specific type of ordering, or specific domains. Following that, the entire  domain may be found through a bond percolation procedure. 

\section{Characterizing self-assembly based on anisotropic building blocks}

\begin{figure}[h!]
    \centering
    \includegraphics[width=0.8\columnwidth]{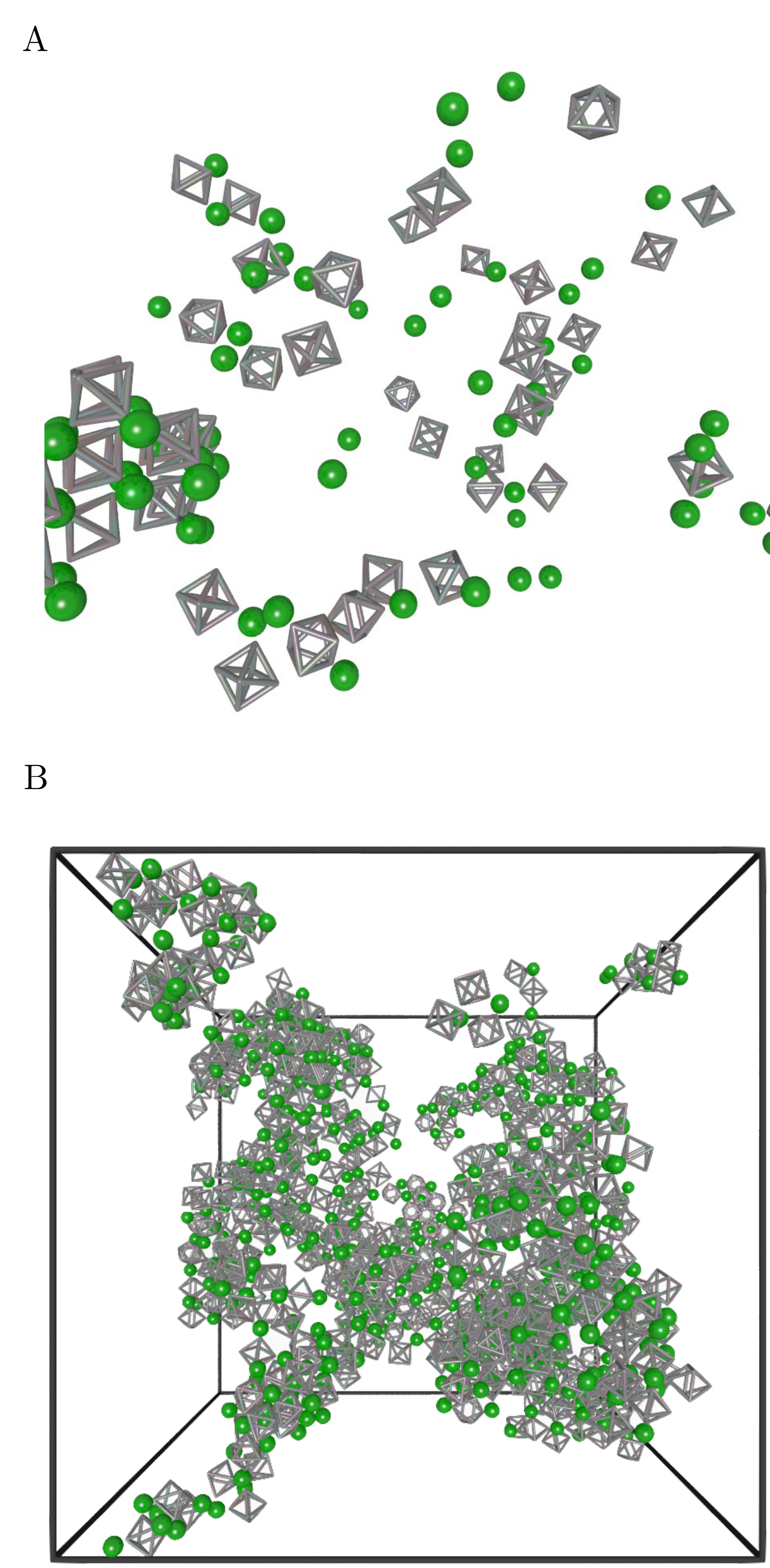}
    \caption{Simulation of self-assembly in  IP/PP hybrid system used to demonstrate the use of SymBOPs. IPs (shown as green) mediate the interaction between the patches of PPs (gray). (A) An intermediate step of the simulation where seeds of assembly begin to form. (B) The final self-assembled state. This configuration is analyzed for domains of orientational order.}
    \label{fig:octahedra_assembly}
\end{figure}

To demonstrate the potential of SymBOPs for characterization of  complex self-assembled structures, here we apply it to a specific example of a hybrid system that contains patchy and isotropic particles (PPs and IPs). In this model, all particles are subject to pairwise hard core repulsion, and the only attractive interaction is between IPs and patches of PPs. This study is motivated by the recent experimental studies of self assembly of designer particles based on DNA origami, combined with DNA-functionalized isotropic nanoparticles \cite{Oleg_hybrid_2016,diamond_oleg}. A detailed discussion of the model and its phenomenology is presented in Ref.~\onlinecite{octahedra}. Here we focus on a specific example of PPs with an octahedral arrangement of patches, to be referred to as Oh-PPs.  An intermediate step of the assembly is shown in Fig.~\ref{fig:octahedra_assembly}A. The analysis can be done at any time  step, but we focus only on the final assembled state shown in Fig.~\ref{fig:octahedra_assembly}B. Within the obtained aggregate, there may be ordered domains  that we would like to identify. There are two types of order we want to investigate: cubatic order associated with relative alignment of the Oh-PPs,   and the bond-orientational order to be characterized with a local SymBOP. 

Because our configuration is obtained through a simulation, and we know the precise orientations of each particle, it is natural to use a SymBOP based on the particle reference. The reference vectors $|R)_l=|s_i)_l$ for  particles with  octahedral (cubic) symmetries are listed in  Table \ref{table:tensors}. The lowest non-vanishing order of the PNOP for this symmetry group is $l=4$, which corresponds to  the traditional cubatic order parameter. According to Eq.~\ref{eq:Sph_harm_BOP_bond}, the magnitude of SymBOP  assigned to a bond between particles $i$ and $j$ is  $(s_i|b_{ij})_l$. Like traditional BOPs, this is a scalar. But in contrast, it contains non-trivial information about the relative alignment between the bond and the particle. In fact, since there are two particles involved, there are two scalar values assigned two each bond.

\begin{figure}[h!]
    \centering
    \includegraphics[width=1\columnwidth]{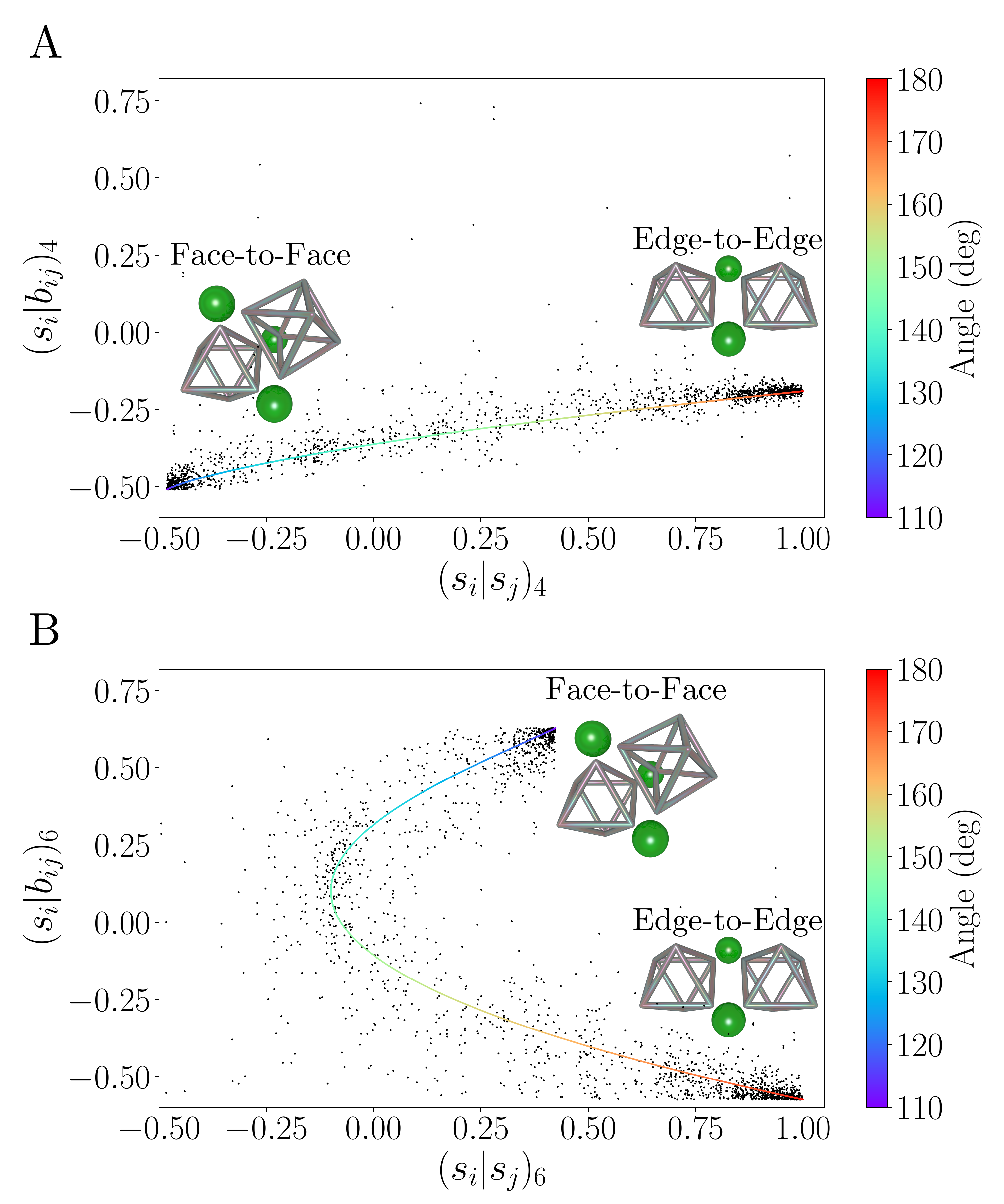}
    \caption{An example of local order parameter scatter plots. Each point represents a single bond. The vertical axis measures the SymBOP, while the horizontal axis is a measure of relative orientational alignment between particles. The color bar shows the angle in degrees between tangents to the nearest faces between the two neighboring PPs. Clusters of bonds for (A) $l=4$ and (B) $l=6$ are found both in the region of high relative orientational order (EtE) and low relative orientational order (FtF), identifying domains of two distinct types of bonding.}
    \label{fig:octahedra_results}
\end{figure}

In addition to SymBOP, the PNOP correlator  $(s_i|s_j)_l$ serves as a measure of the relative  alignment between the  PPs  $i$ and $j$, respectively. The two order parameters can now be combined to find coherent domains within the self-assembled structure. The magnitude of  SymBOP, $(s_i|b_{ij})_l$ can be plotted against the relative alignment of the particles, $(s_i|s_j)_l$, as shown in Fig.~\ref{fig:octahedra_results}, for $l=4,6$. Each  bond  $ij$ is represented by two points in the plot, since $|b_{ij})_l$ can be projected onto two reference vectors, $|s_i)_l$ and $|s_j)_l$, which represent  orientations of each of the two particles. 

% The vertical represents the  magnitude of the SymBOP, and the horizontal axis shows the relative alignment of  the two particles.

% \begin{figure}[h!]
%     \centering
%     \includegraphics[width=1\columnwidth]{Figures/Fig3_SymBOP_plot_2.08.pdf}
%     \caption{An example of local order parameter scatter plots. Each point represents a single bond. The vertical axis measures the SymBOP, while the horizontal axis is a measure of relative orientational alignment between particles. The color bar shows the angle in degrees between tangents to the nearest faces between the two neighboring octahedra. Clusters of bonds for (A) $l=4$ and (B) $l=6$ are found both in the region of high relative orientational order (EtE) and low relative orientational order (FtF), identifying domains of two distinct types of bonding.}
%     \label{fig:octahedra_results}
% \end{figure}

\begin{figure}[h!]
    \centering
    \includegraphics[width=1\columnwidth]{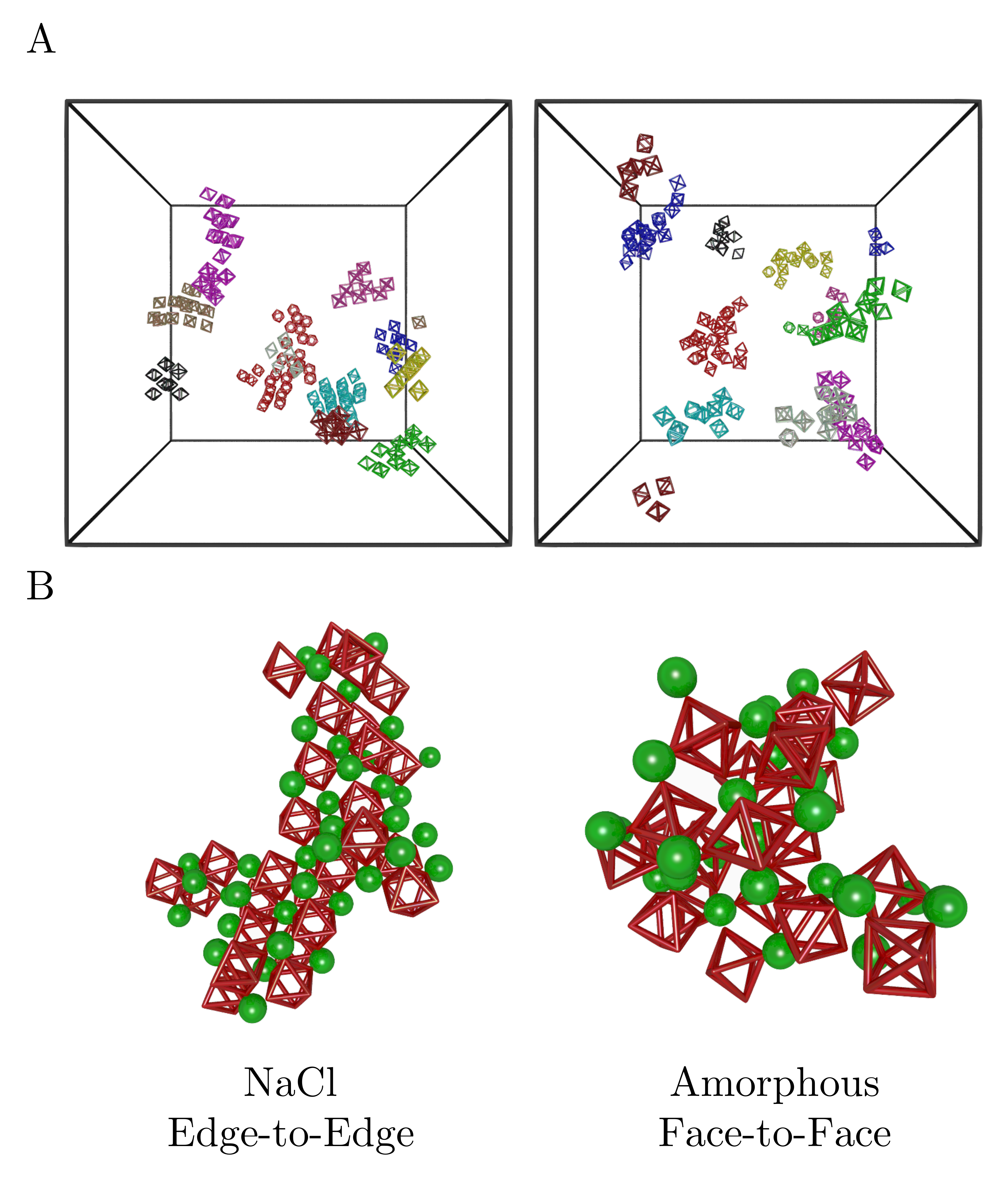}
    \caption{Domains found for the simulation of ideal PPs. (A) shows the domains found for the two types of bonding found in the local order parameter scatter plots. Edge-to-edge bonded domains are shown on the left and face-to-face bonded domains on the right. (B) A closer look at the largest domains of each bonding type found for this simulations state. Long-range order with a NaCl lattice is seen in the edge-to-edge domain, while the face-to-face domains assemble into amorphous structures.}
    \label{fig:octahedra_domains}
\end{figure}

This plot is very informative.  Every type of bond that exists in the sample is represented in this plot and can be individually singled out. For this particular sample we see clusters of bonds near the endpoints of the colored curve shown. This curve, drawn through the points made by the bonds, was found to follow a change in the bonding from edge-to-edge for large values of $(s_i|s_j)_l$ to face-to-face as the insets in the plot show. The color represents the relative angle between the PPs as if it were a hinge on two IPs holding them together. As we move from red to purple, the pair of edge-to-edge PPs fold together until they reach a face-to-face configuration. The limits of these two types of bonds are exactly where we see the clusters of points. This indicates that we are likely to find domains that consist primarily of  edge-to-edge or  face-to-face bonds, respectively. To do this, we keep all of the bonds that fall within a neighborhood centered on the cluster in the plot and use bond percolation to find the domains defined by these ``chosen" bonds. In real space this begins at one bond and branches out through the graph of chosen bonds until it reaches a natural boundary. At this point, if there are chosen bonds remaining, bond percolation is repeated, finding new domains, until all of such bonds are used. Fig.~\ref{fig:octahedra_domains}A shows the domains found in this sample with edge-to-edge bonding on the left and face-to-face bonding on the right. Figure~\ref{fig:octahedra_domains}B is a closer look at the largest domains found of each type with their nanoparticles in green.

The domains with edge-to-edge binding correspond to IPs and PPs arranged into NaCl crystal lattice. Each of these domains is characterized by both long rang cubatic and bond orientation order, i.e. non vanishing PNOP  $\langle |s_i)_l\rangle$ and SymBOP $\langle |s_i)_l (s_i|b_{ij})_l \rangle$, respectively.    In contrast, face-to-face binding results in a peculiar amorphous structure. While this structure lacks a long-range order and cannot fill the space without defects, it is easily identifiable by bond clustering in our scatter plots Fig.~\ref{fig:octahedra_results}. Furthermore, unusual for amorphous aggregates, we were able to identify coherent individual clusters of that structure. That would not be possible with PNOPs or traditional BOPs.  

\section{\label{sec:Conclusion} Conclusion}

As more complex building blocks are introduced for the purposes of self-assembly, more sophisticated analysis methods must be developed. In this paper, we have expanded the use of tensor BOPs to account for the specific symmetry of the particles and/or the  underlying structure. Typically, only rotationally invariant  BOPs are used, which  is limited to a measure of the local degree of crystallinity and general classification of the lattice type in the neighborhood of an isotropic particle. SymBOPs introduced in this work have additional fidelity thanks since they take advantage of anisotropy of the constituent particles.
% Often the traditional method will have difficulties stemming e.g. from an incomplete set of neighbors, especially near the boundaries of a domain.
In effect, we constructed as many images of each bond, as dictated by the underlying symmetry group. As a result, a non trivial value of SymBOP could be assigned already at a single-bond level. This is leveraged by an additional information about relative alignment of the neighboring particles, as measured by PNOP correlator. Once the groups of bonds of a particular type are identified within a structure, the bond percolation procedure is employed to find coherent domains.  

% % of and using the PNOP to symmetrize a BOP, we complete the orbit of each bond according to the symmetry group of the reference tensor.
% % This eliminates the difficulties arising from not having a complete set of neighbors and allows for much better analysis overall. 
% Pairing  SymBOP with a measure of the degree of orientational alignment between the particles themselves allows us to identify two kinds of orientational order: specific orientations of bonds in particle's reference, and  PNOP correlations between neighboring constituent particles.

We have demonstrated the proof-of-concept application of our  method  to simulations of self-assembly in a hybrid system containing patchy and isotropic  particles. The local order parameter scatter plot, seen in Fig.~\ref{fig:octahedra_results}, allows us to identify not only regions with long-range order, such as NaCl lattice with edge-to-edge binding of PPs, but also highly coherent amorphous clusters, such as those obtained by face-to-face binding. Our method is potentially applicable not only to simulations of anisotropic particles but also experiments where the orientations of the particles may be unknown. In the latter case, the self-consistent reference should be used to identify individual domains. This important extension of the method will be addressed in our future studies.

\begin{acknowledgments}
This research was partially done at, and used
resources of the Center for Functional Nanomaterials, which is a U.S.
DOE Office of Science Facility, at Brookhaven National Laboratory under
Contract No.~DE-SC0012704. M.D. ackowledges financial support from NSF CAREER award DMR-1654325. S.C.M. acknowledges financial support from NSF award OAC-1547580 and the Chemical and Biochemical Engineering Department at Rutgers.
\end{acknowledgments}

%\pagebreak

\bibliography{main.bib}

\end{document}